\documentclass[aps,nofootinbib,showpacs,superscriptaddress,preprintnumbers]{revtex4}
\usepackage{graphicx}
\newcommand{\pipsq}[1]{\left[ #1 \right]} 
\newcommand{\pib}[1]{\left[ #1 \right]}

\newcommand{\slashp}{\mbox{$\not \hspace*{-1.10mm} p$}}

\newcommand{\eff}{\mbox{\rm\scriptsize eff}}
\newcommand{\pert}{\mbox{\rm\scriptsize pert}}
\newcommand{\Npert}{\mbox{\rm\scriptsize Npert}}
\newcommand{\NSt}{{\mbox{\scriptsize\it NS}}}
\newcommand{\St}{{\mbox{\scriptsize\it S}}}

\begin{document}

\preprint{ZTF-05-02}


\title{
$\eta$ and $\eta'$ mesons and dimension 
2 gluon condensate $\langle A^2 \rangle$ 
}

\author{Dalibor Kekez}
\affiliation{\footnotesize Rudjer Bo\v{s}kovi\'{c} Institute,
         P.O.B. 180, 10002 Zagreb, Croatia}

\author{Dubravko Klabu\v{c}ar\footnote{Senior associate of Abdus Salam ICTP}}
\affiliation{\footnotesize Department of Physics, Faculty of Science,
        Zagreb University, Bijeni\v{c}ka c. 32, 10000 Zagreb, Croatia}


\begin{abstract}
The study of light pseudoscalar quark-antiquark bound states 
in the Dyson-Schwinger approach with the {\it effective} QCD 
coupling enhanced by the interplay of the dimension 2 gluon 
condensate $\langle A^2 \rangle$ and dimension 4 gluon condensate 
$\langle F^2 \rangle$, is extended to the $\eta$--$\eta'$ complex. 
We include the effects of the gluon axial anomaly into the 
Dyson-Schwinger approach to mesons. 
The calculated masses, mixing and two-photon 
decay widths of $\eta$ and $\eta'$ mesons are in agreement
with experiment.
Also, in a model-independent way, 
we give the modification of the Gell-Mann--Okubo and 
Schwinger nonet relations due to the interplay of 
the gluon anomaly and SU(3) flavor symmetry breaking.
\end{abstract}

\maketitle

\section{Introduction}
\label{INTRO}

The dimension-2 gluon condensate 
$\langle A_\mu^a A^{a\mu}\rangle \equiv \langle A^2 \rangle$
attracted the attention of some researchers well over a decade ago; 
e.g., see Refs. 
\cite{Celenza:1986th,Lavelle:eg,Lavelle:xg,Ahlbach:ws,Lavelle:yh}.
However, there was a wide-spread opinion that, since this condensate
is not gauge invariant, it cannot have observable consequences
and cannot play an important role in QCD. 
In contrast to that, the gauge-invariant, dimension-4 gluon condensate
$\langle F_{\mu\nu}^a F^{a\mu\nu} \rangle \equiv \langle F^2 \rangle$
\cite{Shifman:bx} has, over 25 years now, been a subject of studies 
discussing its value (e.g., see Ref. \cite{Ioffe:2002be})
and implications for QCD.

After it turned out more recently that the Landau-gauge value of 
$\langle A^2 \rangle$ corresponds to a more general gauge-invariant
quantity, it attracted a lot of theoretical attention
\cite{Boucaud:2000nd,Gubarev:2000eu,Gubarev:2000nz,Kondo:2001nq,Kondo:2001tm,Boucaud:2002fx,Dudal:2002xe,Kekez:2003ri,Slavnov:2004rz,RuizArriola:2004en,Li:2004zu},
to quote just several of many papers offering evidence that 
$\langle A^2 \rangle$ condensate may be important for the 
nonperturbative regime of Yang-Mills theories.  In  
Ref. \cite{Kekez:2003ri}, we argued that $\langle A^2 \rangle$ 
condensate may be relevant for the Dyson-Schwinger (DS) approach 
to QCD. Namely, in order that this 
approach \cite{reviews,Alkofer:2000wg,Roberts:2000aa,Maris:2003vk} 
leads to a successful hadronic phenomenology [which has so far been 
treated {\it widely} only in the rainbow-ladder approximation (RLA)], 
an enhancement of the effective quark-gluon interaction seems to be 
needed at intermediate spacelike momenta, $Q^2 \sim 0.5$ GeV$^2$.
Reference \cite{Kekez:2003ri} showed that the interplay of the
dimension-2 condensate $\langle A^2 \rangle$ with the dimension-4 
condensate $\langle F^2 \rangle$ can provide such an enhancement. 
It also showed that the resulting effective strong running 
coupling leads to the sufficiently strong dynamical chiral 
symmetry breaking (D$\chi$SB) and successful phenomenology 
in the sector of light pseudoscalar mesons. In addition, 
the issues such as quark propagator solutions, $p^2$-dependent
dressed (``constituent'') quark masses and a more detailed
discussion of the parameter dependence of the results were
addressed in Refs. \cite{Kekez:2004tm,Bled2003}. 

In the present paper, we extend the treatment of the 
nonzero-isospin light pseudoscalar mesons of 
Ref. \cite{Kekez:2003ri} to the $\eta$--$\eta'$ complex.
First, in the next section, the key result of Ref. \cite{Kekez:2003ri}, 
namely its gluon-condensate-enhanced interaction (\ref{ourAlpha_eff}), 
is briefly re-derived in another, less rigorous and heuristic way.
In the third section, we review how the DS approach employing 
such an interaction, can
give the successful light (i.e., involving quark flavors $q=u,d,s$)
meson phenomenology, as this is also needed for the good description of 
the presently interesting $\eta$ and $\eta'$ mesons. Nevertheless, 
for $\eta$ and $\eta'$ this is not enough because of the influence
of the gluon anomaly and thus we explain how its effects are 
included in the manner of Refs. \cite{Klabucar:1997zi,Kekez:2000aw}.
The implementation of the anomaly and the SU(3) flavor symmetry 
breaking, as well as their interplay, are presented independently of 
any concrete dynamics in subsection \ref{etaMasses}, and in detail, 
because of some renewed interest in the flavor dependence of the 
mixing in the $\eta$--$\eta'$ complex (e.g., Ref. \cite{Li:2002hq} 
and references therein).
After that, the masses, mixing angle and two-photon ($\gamma\gamma$) 
decay widths of $\eta$ and $\eta'$ are calculated. Discussion and 
conclusion are in Sec. \ref{last}.

\section{Strong coupling enhanced by gluon condensates}
\label{heuristic}

Reference \cite{Kekez:2003ri} showed how the interaction (\ref{ourAlpha_eff}),
phenomenologically successful in DS studies in the Landau gauge and 
RLA, resulted from combining the form [Eq.  (\ref{Alkofalpha}) below]
which the running coupling has in the Landau-gauge DS studies
\cite{Alkofer:2000wg,Alkofer:2002ne,Fischer:2003rp,Alkofer:2003vj,Bloch:2003yu}
and the ideas on the possible relevance of the $\langle A^2 \rangle$
gluon condensate
\cite{Boucaud:2000nd,Gubarev:2000eu,Gubarev:2000nz,Kondo:2001nq,Kondo:2001tm,Boucaud:2002fx,Dudal:2002xe,Lavelle:eg,Lavelle:xg,Ahlbach:ws,Lavelle:yh}.
In the present paper, we give a simplified and
more intuitive derivation thereof as follows.

The full gluon propagator $D_{\mu\nu}^{ab}(k)$ in the Landau gauge is 
defined through the free gluon propagator $D_{\mu\nu}^{ab}(k)_{0}$
and the gluon renormalization function $Z(-k^2)$ like this:
\begin{equation}
D_{\mu\nu}^{ab}(k) = Z(-k^2) D_{\mu\nu}^{ab}(k)_{0} \equiv 
 Z(-k^2) \,\, \frac{\delta^{ab}}{k^2} 
\left( -g_{\mu\nu} + \frac{k_\mu k_\nu}{k^2} \right)
\, .
\label{gluonLGpropag}
\end{equation}
The full ghost propagator $D_G(k)$ is similarly defined by 
the ghost renormalization function $G(-k^2)$:
\begin{equation}
D_G(k) = \frac{ G(-k^2)}{k^2} \, .
\label{ghostLGpropag}
\end{equation}
The strong running coupling $\alpha_{\mbox{\rm\scriptsize s}}(Q^2)$
used in the Landau-gauge DS studies
\cite{Alkofer:2000wg,Alkofer:2002ne,Fischer:2003rp,Alkofer:2003vj,Bloch:2003yu}
is defined as
\begin{equation}
\alpha_{\mbox{\rm\scriptsize s}}(Q^2)
=
\alpha_{\mbox{\rm\scriptsize s}}(\mu^2) \, Z(Q^2) \, G(Q^2)^2
\, ,
\label{Alkofalpha}
\end{equation}
where $\alpha_{\mbox{\rm\scriptsize s}}(\mu^2) = g^2/4\pi$ and
$Z(\mu^2) G(\mu^2)^2 = 1$ at the renormalization point $Q^2 = \mu^2$.
Our convention is $k^2 = -Q^2 < 0$ for spacelike momenta $k$.

The functions $Z$ and $G$ can be 
expressed through the corresponding gluon ($A$) and ghost ($G$) 
polarization functions $\Pi_A(Q^2)$ and $\Pi_G(Q^2)$:
\begin{equation}
Z(Q^2) = \frac{1}{1+\frac{\Pi_A(Q^2)}{Q^2}} \,  
\qquad   ,  \qquad
G(Q^2) = \frac{1}{1+\frac{\Pi_G(Q^2)}{Q^2}} \, .
\label{A+Gren}
\end{equation}
Almost two decades ago, it was noted that in the operator product 
expansion (OPE) the gluon condensate $\langle A^2 \rangle$ can 
contribute to QCD propagators; e.g., see Refs.
\cite{Lavelle:eg,Lavelle:xg,Ahlbach:ws,Lavelle:yh}.
Their $\langle A^2 \rangle$-contributions to the OPE-improved
gluon ($A$) and ghost ($G$) polarization functions were 
more recently confirmed by Kondo \cite{Kondo:2001nq}.
For the Landau gauge adopted throughout this paper,
three QCD colors ($N_c = 3$) and four space-time dimensions 
($D=4$), their expressions for the polarizations become
\begin{equation}
\Pi_i(Q^2) = m_i^2 + {\cal O}_i(1/Q^2) \, ,
\qquad (i=A,G)
\, ,
\label{Kondo}
\end{equation}
\begin{equation}
m_A^2 = \frac{3}{32} \,\,  g^2 \langle A^2 \rangle = - m_G^2~,
\label{gluonMass}
\end{equation}
where $m_A$ and $m_G$ are, respectively, dynamically generated
effective gluon and ghost mass. 
Reference \cite{Kekez:2003ri} estimated $m_A = 0.845$ GeV (and found 
that it was, phenomenologically, a remarkably successful initial 
estimate) by using in Eq. (\ref{gluonMass}) the lattice result 
$g^2 \langle A^2 \rangle = 2.76$ GeV$^2$ of Ref. \cite{Boucaud:2000nd},
a value compatible with the bound resulting from the discussions of
Gubarev {\it et al.} \cite{Gubarev:2000eu,Gubarev:2000nz} on the
physical meaning of $\langle A^2 \rangle$
and its possible importance for confinement.

As for the contributions ${\cal O}_i(1/Q^2)$ ($i=A,G$) in Eq. 
(\ref{Kondo}), it turned out \cite{Lavelle:xg,Ahlbach:ws,Lavelle:yh}
that they contain many kinds of mostly unknown 
condensates [e.g., gauge-dependent gluon, ghost and mixed ones, 
where terms $\propto (1/Q^2)^n$ $(n>1)$ were not considered at all].
The only practical approach at this point is therefore that these
complicated contributions 
are approximated by the terms $\propto 1/Q^2$ and parameterized, i.e.,
\begin{equation}
{\cal O}_A(1/Q^2) \approx \frac{C_A}{Q^2} \,\,\,  , \qquad
{\cal O}_G(1/Q^2) \approx  \frac{C_G}{Q^2} \,\,\,  .
\label{calOs}
\end{equation}
Thus, $C_A$ and $C_G$ would in principle be free parameters 
to be fixed by phenomenology. However, as noted in Ref. 
\cite{Kekez:2003ri}, in the effective gluon propagator proposed 
by Lavelle \cite{Lavelle:ve}, the ${\cal O}_A(1/Q^2)$ polarization
is (for the Landau gauge and $D=4$) given by the dimension-4 gluon
condensate $\langle F^2 \rangle$ as 
\begin{equation}
\Pi_A^{\langle F^2 \rangle}(Q^2)
      =
 \frac{34 N_c \pi \alpha_s \langle F^2 \rangle}{9(N_c^2 - 1)Q^2}
      =
 \frac{(0.640 \,\, {\rm GeV})^4}{Q^2} \, .
\label{PiLavelle91}
\end{equation}
Since Lavelle's \cite{Lavelle:ve} propagator misses some unknown 
three- and four-gluon contributions \cite{Ahlbach:ws,Lavelle:yh}
and since the precise value of $\alpha_s \langle F^2 \rangle$ is 
still not certain, we regard the value 
$C_A = (0.640 \,\, {\rm GeV})^4$ just as an inspired initial 
estimate. Still, together with the assumption $C_G = C_A$, 
it was a very useful starting guess in our 
Refs. \cite{Kekez:2003ri,Kekez:2004tm,Bled2003}, 
leading to very good phenomenological fits.

We are now prepared to give a general, although heuristic argument why the 
contribution (\ref{gluonMass}) of the dimension-2 $\langle A^2 \rangle$ 
condensate to the gluon and ghost polarization functions (\ref{Kondo}),
should indeed lead to the form of $\alpha_{\eff}(Q^2)$ already found
through a more detailed argument in Ref. \cite{Kekez:2003ri}. 
Our first step is to assume that in the gluon and ghost polarizations
(\ref{Kondo}), $\Pi_{A}$ and $\Pi_{G}$, one can disentangle the
perturbative ($\pert$) from nonperturbative ($\Npert$) parts,
$\Pi_i = \Pi^{\pert}_i + \Pi^{\Npert}_i$ $(i=A,G)$. At least for high
momenta $Q^2$, it is then possible to approximately factor away
the perturbative from nonperturbative contributions;
for $i=A$,
\begin{equation}
Z(Q^2)
\approx
\frac{1}{1+\frac{\Pi_A^{\pert}(Q^2)}{Q^2}}    \,
\frac{1}{1+\frac{\Pi_A^{\Npert}(Q^2)}{Q^2}}
  \equiv  Z^{\pert}(Q^2) \, Z^{\Npert}(Q^2)
\, ,
\label{Zapprox}
\end{equation}
where the approximation means neglecting the contribution of
the term
\begin{equation}
\frac{\Pi_A^{\pert}(Q^2)\Pi_A^{\Npert}(Q^2)}{Q^4}  \, .
\label{neglected}
\end{equation}
Analogously,  
\begin{equation}
G(Q^2)
\approx
\frac{1}{1+\frac{\Pi_G^{\pert}(Q^2)}{Q^2}}    \,
\frac{1}{1+\frac{\Pi_G^{\Npert}(Q^2)}{Q^2}}
  \equiv  G^{\pert}(Q^2) \, G^{\Npert}(Q^2)
\, .
\label{Gapprox}
\end{equation}

Since the general QCD coupling 
$\alpha_{\mbox{\rm\scriptsize s}}(Q^2)$ 
must reduce to the {\it perturbative} QCD coupling 
$\alpha_{\pert}(Q^2)$ for so very high $Q^2$ that
nonperturbative contributions are negligible,
Eq. (\ref{Alkofalpha}) implies that
\begin{equation}
\alpha_{\mbox{\rm\scriptsize s}}(\mu^2) \, 
Z^{\pert}(Q^2) \, G^{\pert}(Q^2)^2 
= \alpha_{\pert}(Q^2) \, .
\label{alpha_pert}
\end{equation}
We can also assume, for high $Q^2$, that nonperturbative parts are
given by the OPE-based results of Refs. 
\cite{Lavelle:eg,Lavelle:xg,Ahlbach:ws,Lavelle:yh,Kondo:2001nq}. In 
our present case, they amount to Eqs. (\ref{Kondo})-(\ref{gluonMass})
for the gluon and ghost polarizations, and to the parameterization 
(\ref{calOs}). This then gives
\begin{eqnarray}
Z^{\Npert}(Q^2)&=&\frac{1}{1
                   +\frac{m_A^2 }{Q^2}
                   + \frac{C_A}{Q^4}}~,
\label{ZOPE}
\\
G^{\Npert}(Q^2)&=&\frac{1}{1
                   -\frac{m_A^2}{Q^2}
                   + \frac{C_G}{Q^4}}~.
\label{ZGOPE}
\end{eqnarray}
Equations (\ref{Alkofalpha}), (\ref{Zapprox}), (\ref{Gapprox}), 
(\ref{ZOPE}), and (\ref{ZGOPE}), considered together, then 
suggest an effective coupling $\alpha_{\eff}(Q^2)$ of the form
\begin{equation}
\alpha_{\eff}(Q^2)
 =
\alpha_{\pert}(Q^2) \,
Z^{\Npert}(Q^2) \, G^{\Npert}(Q^2)^2~.
\label{ourAlpha_eff}
\end{equation}
                  
Obviously, the above derivation of the coupling (\ref{ourAlpha_eff}) 
is only heuristic, but we have already 
presented its more rigorous derivation in Ref. \cite{Kekez:2003ri}.

In Refs. \cite{Kekez:2003ri, Kekez:2004tm, Bled2003},
we discussed why, how and when the form (\ref{ourAlpha_eff}) was 
sufficiently enhanced at intermediate $Q^2$ to lead to the
successful pion and kaon phenomenology when used in the 
DS approach in RLA.

\section{DS approach and its extension to \mbox{$\eta$--$\eta^\prime$}
complex}
\label{MainSec}

In the DS approach to QCD, one solves the gap equation, i.e., 
the DS equation for quark two-point functions, namely
dressed quark propagators 
\begin{equation}
S_q(p) \, = \,
\frac{1}{ \slashp \, {\cal A}_q(p^2) - {\cal B}_q(p^2) }
  \, \equiv \,
\frac{ {\cal A}_q(p^2)^{-1}}{ \slashp  - {\cal M}_q(p^2) }
\label{FULLqPropag}
\end{equation}
of various flavors $q$, and so explicitly constructs 
constituent quarks characterized by the dynamical masses 
${\cal M}_q(p^2) \equiv {\cal B}_q(p^2)/{\cal A}_q(p^2)$. 
The constituent quarks and antiquarks of respective flavors
$q$ and $q'$ build meson bound states, which are 
solutions of the Bethe-Salpeter (BS) equation for the 
bound-state vertex $\Gamma_{q{\bar q}'}$:
\begin{equation}
[\Gamma_{q{\bar q}'}(k,{P})]_{ef} = \int \frac{d^4\ell}{(2\pi)^4}
[S_q(\ell+\frac{{P}}{2}) \Gamma_{q{\bar q}'}(\ell,{P})
S_{q'}(\ell-\frac{{P}}{2}) ]_{gh} [K(k-\ell)]_{ef}^{hg}~,
\label{BSE}
\end{equation}
where $e,f,g,h$ {\it schematically} represent spinor, color
and flavor indices. Solving Eq. (\ref{BSE}) for $\Gamma_{q{\bar q}'}$
also yields $M_{q\bar q'}$, the mass eigenvalue
of the $q{\bar q}'$ bound state. 
Unfortunately, the {\it full} interaction kernel $K(k,\ell,{P})$ 
for the BS equation is not known.
Also, the full kernels of the gap equations [which supply 
the quark propagators $S_q(\ell)$ (\ref{FULLqPropag})
to Eq. (\ref{BSE})] involve the full gluon propagator and 
the full quark-gluon vertex,
which satisfy their own DS equations. They in turn involve 
higher $n$-point functions and their DS equations, etc. 
This infinite tower of the integral DS equations must be truncated
to make the problem tractable 
\cite{reviews,Alkofer:2000wg,Roberts:2000aa,Maris:2003vk}. 
The approximations employed
in the gap equation and the BS equation must be mutually consistent
in order to preserve the important characteristics of the full theory.
In the low-energy sector of QCD, the nonperturbative phenomenon of 
D$\chi$SB is the most important feature.  Phenomenological DS studies 
have therefore mostly been relying on the {\it consistently} used 
RLA, where D$\chi$SB is well understood
\cite{Alkofer:2000wg,Maris:2003vk,Jain:1993qh,Maris:1997tm,Maris:1999nt}. 
The consistent RLA implies that for interactions between quarks 
one uses {\it Ans\" atze} of the form      
\begin{equation}
[K(k)]_{ef}^{hg} = {\rm i} \, 
4\pi\alpha_{\mbox{\rm\scriptsize eff}}(-k^2) \,
       D_{\mu\nu}^{ab}(k)_{0} \,
[\frac{\lambda^a}{2}\,\gamma^{\mu}]_{eg} \,
[\frac{\lambda^a}{2}\,\gamma^{\nu}]_{hf} \, 
\label{RLAkernel}
\end{equation}
both in the BS equation (\ref{BSE}) and the gap equation 
(\ref{DS-equation}) for the dressed quark propagators (\ref{FULLqPropag}),
\begin{equation}
        S_q^{-1}(p) = \slashp - \widetilde{m}_q - \, {\rm i} \,
4\pi \int \!\frac{d^4\ell}{(2\pi)^4} \, \alpha_{\eff}[-(p-\ell)^2]
D_{\mu\nu}^{ab}(p-\ell)_{0} \, \frac{\lambda^a}{2}\,\gamma^{\mu}
S_q(\ell) \frac{\lambda^b}{2}\,\gamma^{\nu}~.
       \label{DS-equation}
        \end{equation}
Here, $\widetilde{m}_q$ is the bare mass of the quark flavor $q$
breaking the chiral symmetry explicitly.
The case $\widetilde{m}_q=0$ corresponds to the chiral limit where
the current quark mass $m_q=0$.

One solves the DS equation (\ref{DS-equation}) for dressed 
propagators of the light quarks ($q=u,d,s$) using a, hopefully,
phenomenologically successful interaction [in the present paper,
it is given by Eq. (\ref{ourAlpha_eff})]. These light-quark 
propagators are then employed in BS equations 
for quark-antiquark ($q\bar q$) relativistic bound states.

The most important advantage of adopting RLA in the context of 
low-energy QCD is that D$\chi$SB and, consequently, 
the appearance of light pseudoscalar mesons as (almost-)Goldstone 
bosons in (the vicinity of) the chiral limit, is well-understood 
and under control in this approximation scheme 
\cite{Alkofer:2000wg,Maris:2003vk,Jain:1993qh,Maris:1997tm,Maris:1999nt}.
Solving the BS equation in the chiral limit gives 
the vanishing pion mass, $M_\pi=0$.
More generally and precisely, the light pseudoscalar masses 
$M_{q\bar q} \propto \sqrt{{\widetilde m}_{q}}$, as required by QCD
through Gell-Mann--Oakes--Renner (GMOR) relation. 
All this is a manifestation of the correct chiral 
QCD behavior in the DS approach, in which all light pseudoscalar mesons 
($\pi^{0,\pm}, K^{0,\pm}, {\bar{K^0}}, \eta$) manifest 
themselves {\it both} as 
$q\bar q$ bound states 
{\it and} (almost-)Goldstone bosons of dynamically broken chiral symmetry. 
This resolution of the dichotomy ``$q\bar q$ bound state {\it vs.}
Goldstone boson", enables one to work with the mesons as
explicit $q\bar q$ bound states 
while reproducing (even analytically in the chiral limit) the famous
results of the axial anomaly for the light pseudoscalar mesons, 
most notably the $\pi^0 \to \gamma\gamma$ decay amplitude
$T_{\pi^0}^{\gamma\gamma} = 1/(4\pi^2 f_\pi)$ 
in the chiral limit \cite{Roberts:1994hh,Bando:1993qy}.
This is unique among the bound state approaches -- e.g., see
Refs. \cite{Roberts:2000hi,Kekez:1998xr} and references
therein for discussion. Nevertheless, one keeps the advantage
of bound state approaches that from the $q\bar q$ substructure
one can calculate many important quantities (such as the pion decay
constant $f_\pi$) which are just parameters in most of other chiral
approaches to the light-quark sector.

\begin{table}
\caption{ 
Masses $M_P$, decay constants $f_P$ and $\gamma\gamma$
decay amplitudes $T_{P}^{\gamma\gamma}$ of the pseudoscalar
$q{\bar q}'$ bound states $P=\pi, K$ and $s\bar s$
resulting from our gluon condensate-induced
$\alpha_{\mbox{\rm\scriptsize eff}}(Q^2)$ (\ref{ourAlpha_eff}),
along with the parameter values fixed in Ref. \cite{Kekez:2003ri}
by fitting the pion and kaon properties -- see
Eq. (\ref{StandardParameterSet-new}) and the text below it.
These masses and decay constants are the input for
the description of the $\eta$--$\eta^\prime$ complex.
The last column is the constituent quark mass ${\cal M}_q(0)$
pertinent to the corresponding $q\bar q$ meson,
i.e., ${\cal M}_u(0)={\cal M}_d(0)$ for $P=\pi$ and
${\cal M}_s(0)$ for $P=s\bar s$. [Later, in Eq. (\ref{etaSdef}),
we will name the unphysical $s\bar s$ pseudoscalar meson
$\eta_\St$, but note that the mass $M_{\eta_\St}$ (\ref{MetaS}),
introduced in the {\it NS--S} mass matrix (\ref{M2_NS-S}), 
includes the contribution from the gluon anomaly,
whereas $M_{s\bar s}$ does not.]
}
\begin{ruledtabular}
\begin{tabular}{cccccc}
  $ P $      & $M_{P}$ & $f_{P}$ & $1/(4\pi^2 f_{P})$ & $T_{P}^{\gamma\gamma}$ & ${\cal M}_q(0)$ \\
\hline
  $\pi$     & {0.1350  } & {0.09293  } & {0.2726  }       & {0.2560  }  & {0.3842  } \\
\hline
  $K$       & {0.4949  } & {0.1115  }  &                       &                  &                 \\
\hline
  $s\bar s$ & {0.7221  } & {0.1329  }  & {0.1905  }  & {0.08599  } & {0.5922  } \\
\end{tabular}
\end{ruledtabular}
\label{piKssbarTable}
\end{table}

In the DS approach, the D$\chi$SB is obtained in an, essentially, 
Nambu--Jona-Lasinio fashion, but the DS model interaction, encoded in 
various popular forms of $\alpha_{\eff}(p^2)$, is usually less schematic. 
Typically, as in Refs. 
\cite{Jain:1993qh,Klabucar:1997zi,Kekez:1998xr,Kekez:1996az,Kekez:1998rw,Roberts:2000aa} 
for example, it combines the modeled nonperturbative component 
(which is strong enough to obtain required D$\chi$SB)
and the known perturbative QCD contribution.
Our gluon condensate-induced 
$\alpha_{\mbox{\rm\scriptsize eff}}(Q^2)$ (\ref{ourAlpha_eff})
also behaves nonperturbatively at low $Q^2$ and
perturbatively at very high $Q^2$,
as seen both in its earlier derivation \cite{Kekez:2003ri}
and the alternative derivation in Sec. \ref{heuristic}. 
Thus, as shown in 
Ref. \cite{Kekez:2003ri}, our interaction (\ref{ourAlpha_eff}) leads 
to the deep Euclidean behaviors of quark propagators 
consistent with the asymptotic freedom of QCD \cite{Kekez:1998xr}.
However, what enables the successful description of pions and kaons
in Ref. \cite{Kekez:2003ri} and what is also crucial in the present 
paper, are the contributions to the gap and BS equations at low and 
intermediate momenta.
The crucial result is the behavior of the gap equation solutions 
for the mass functions ${\cal M}_q(Q^2)$ at 
$Q^2=0$ to $Q^2\approx (500 \, {\rm MeV})^2$, where ${\cal M}_q(Q^2)$
($q=u,d,s)$,
due to D$\chi$SB, have values consistent with typical values
of the constituent mass parameters in constituent quark models.
[When we need to be specific, we can, for definiteness, choose to 
call the $Q^2=0$ value ${\cal M}_q(0)$ the constituent mass.]
In the chiral limit, where, as usual in the consistent DS approach 
\cite{reviews,Alkofer:2000wg,Roberts:2000aa,Maris:2003vk},
we get the correct chiral Goldstone-pion behavior, 
our $\alpha_{\mbox{\rm\scriptsize eff}}(Q^2)$ (\ref{ourAlpha_eff}) 
with the parameters from Ref. \cite{Kekez:2003ri}, namely 
\begin{equation}
C_A = (0.6060 \,\, {\rm GeV})^4 = C_G \,\, , \,\,
m_A = 0.8402 \,\, {\rm GeV} \, ,
\label{StandardParameterSet-new}
\end{equation}
gives us ${\cal M}_{u,d}(0)=369$ MeV, and empirically already quite 
acceptable values for the pion decay constant and
the $\langle {\bar q} q \rangle$ condensate. 
Away from the chiral limit (but still 
for isosymmetric $u$- and $d$-quarks),
the same interaction
yields ${\cal M}_{u,d}(0)=384$ MeV [just 4\% above 
${\cal M}_{u,d}(0)$ in the chiral limit] with the bare mass
${\widetilde m}_{u,d}=3.046$ MeV, for which value our model 
reproduces the experimental values of $\pi^0$ mass, $f_{\pi^+}$ 
decay constant as well as $\pi^0 \to \gamma\gamma$ decay amplitude, 
and respects GMOR relation \cite{Kekez:2003ri}. The empirical 
values of the kaon mass and decay 
constant were also reproduced very well when one in addition takes 
${\widetilde m}_s = 67.70 $ MeV \cite{Kekez:2003ri}. The $s$ quark 
constituent, dynamical mass is then ${\cal M}_s(0)=592$ MeV.  
(In our earlier papers \cite{Kekez:2003ri,Kekez:2004tm,Bled2003}
we also found that the results for ${\widetilde m}_{u,d}$ and 
${\widetilde m}_s$ were rather robust. For example, the
values quoted here because they are preferred when the
interaction (\ref{ourAlpha_eff}) is 
employed in the gap and BS equations, (\ref{DS-equation}) 
and (\ref{BSE}), are quite close to the values
of ${\widetilde m}_{u,d}$ and ${\widetilde m}_s$ preferred
when the Jain--Munczek effective interaction \cite{Jain:1993qh}
is used instead.)

Up to accounting for the gluon anomaly,
the results in Table \ref{piKssbarTable} are the input from
the well-described pion and kaon sector \cite{Kekez:2003ri}
which enables, without any refitting of the model parameters
${\widetilde m}_{u,d}$, ${\widetilde m}_s$ and 
(\ref{StandardParameterSet-new}), the good description of
the $\eta$--$\eta^\prime$ complex.

\subsection{Masses in the \mbox{$\eta$--$\eta^\prime$} complex}
\label{etaMasses}

The description 
\cite{Klabucar:1997zi,Kekez:2000aw,Kekez:2001ph}
of $\eta$ and $\eta'$ is especially noteworthy, as it is 
successful in spite of the limitations of the DS approach
in the ladder approximation.
For this description, the crucial issues are the 
meson mixing and construction of physical meson states. For the DS 
approach, they are formulated in Refs. \cite{Klabucar:1997zi,Kekez:2000aw},
where solving of appropriate BS equations (\ref{BSE})
yields the eigenvalues of the squared masses,
$M_{u\bar{u}}^2,M_{d\bar{d}}^2,M_{s\bar{s}}^2$, and $M_{u\bar{s}}^2$,
of the respective quark-antiquark bound states $|u\bar{u}\rangle , 
|d\bar{d}\rangle , |s\bar{s}\rangle$, and $|u\bar{s}\rangle$.
The last one is simply the kaon, and $M_{u\bar{s}}$ is its mass $M_K$.
Nevertheless, the first three do not correspond to any physical
pseudoscalar mesons.
Thus, $M_{u\bar{u}}^2,M_{d\bar{d}}^2,M_{s\bar{s}}^2$ do not 
automatically represent any physical masses, although the mass matrix
(to be precise, its {\it non-anomalous part}, which vanishes in the
chiral limit) is simply 
\begin{equation}
{\hat M}^2_{NA} = 
\left[ \begin{array}{ccl} M_{u\bar{u}}^2 & \, 0 & \, 0 \\
                          0 & \, M_{d\bar{d}}^2 & \, 0 \\
                          0 & \, 0 & M_{s\bar{s}}^2 \end{array} \right]
\label{diagM2NA}
\end{equation}
in the basis $|q\bar{q}\rangle, (q=u,d,s)$. However, the flavor SU(3) quark
model, and especially the almost exact isospin symmetry, leads one to recouple 
these states into the familiar SU(3) octet-singlet basis of the zero-charge 
subspace of the light unflavored pseudoscalar mesons of well-defined isospin 
quantum numbers:
\begin{eqnarray}
        |\pi^0\rangle
        &=&
        \frac{1}{\sqrt{2}} (|u\bar{u}\rangle - |d\bar{d}\rangle)~,
\label{pi0def}
        \\
        |\eta_8\rangle
        &=&
        \frac{1}{\sqrt{6}} (|u\bar{u}\rangle + |d\bar{d}\rangle
                                            -2 |s\bar{s}\rangle)~,
\label{eta8def}
        \\
        |\eta_0\rangle
        &=&
        \frac{1}{\sqrt{3}} (|u\bar{u}\rangle + |d\bar{d}\rangle
                                             + |s\bar{s}\rangle)~.
\label{eta0def}
        \end{eqnarray}
With $|u\bar{u}\rangle$ and $|d\bar{d}\rangle$ being practically
chiral states as opposed to a significantly heavier $|s\bar{s}\rangle$,
Eqs.~(\ref{pi0def})--(\ref{eta0def}) do not define the octet and singlet 
states of the exact SU(3) flavor symmetry, but the {\it effective} octet 
and singlet states. However, 
in spite of this flavor symmetry breaking by the $s$ quark, these equations 
still implicitly assume nonet symmetry in the sense \cite{Gilman:1987ax} 
that the same states $|q\bar{q}\rangle$ ($q=u,d,s$)
appear in both the octet member $\eta_8$ (\ref{eta8def}) and the singlet
$\eta_0$ (\ref{eta0def}).
Nevertheless, in order to avoid the U$_{\rm A}$(1) problem, 
this symmetry must
ultimately be broken by gluon anomaly at least at the level of the
masses of pseudoscalar mesons.

In the basis (\ref{pi0def})--(\ref{eta0def}), the non-anomalous 
part of the mass matrix of $\pi^0$ and etas is \begin{equation}
{\hat M}^2_{NA} =
\left[ \begin{array}{ccl} M_{\pi}^2 & 0 & 0 \\
                          0 & M_{88}^2 & M_{80}^2\\
                          0 & M_{08}^2 & M_{00}^2
        \end{array} \right]~,
\label{M2NA}
\end{equation}
showing that the isospin $I=1$ state $\pi^0$ does not mix
with the $I=0$ states $\eta_8$ and $\eta_0$, 
thanks to our working in the isospin limit, where
$M_{u\bar{u}} = M_{d\bar{d}}$, which we then can strictly identify 
with our model pion mass $M_{\pi}$. Since in this model we can 
also calculate
$M_{s\bar{s}}^2 = \langle s\bar{s} | {\hat M}^2_{NA} | s\bar{s} \rangle$,
this gives us our calculated entries in the mass matrix:
\begin{equation}
M_{88}^2 \equiv \langle \eta_8 | {\hat M}^2_{NA} |\eta_8 \rangle
= \frac{2}{3}\, (M_{s\bar{s}}^2 + \frac{1}{2}M_{\pi}^2)
= (594.72 \,\, {\rm MeV})^2 \, ,
\label{M88}
\end{equation}
\begin{equation}
M_{80}^2 \equiv \langle \eta_8 | {\hat M}^2_{NA} |\eta_0 \rangle
= M_{08}^2 = \frac{\sqrt{2}}{3} \, ( M_{\pi}^2 - M_{s\bar{s}}^2 )
= - (487.05 \,\, {\rm MeV})^2  \, ,
\label{M80}
\end{equation}
\begin{equation}
M_{00}^2 \equiv \langle \eta_0| {\hat M}^2_{NA} |\eta_0 \rangle
= \frac{2}{3} \, (\frac{1}{2}M_{s\bar{s}}^2 + M_{\pi}^2) 
= (431.22 \,\, {\rm MeV})^2 \, .
\label{M00}
\end{equation}
The values on the far right of these equations were calculated from 
$M_{\pi}=M_{u\bar{u}}$ and $M_{s\bar{s}}$ from Table \ref{piKssbarTable},
i.e., they result from the parameters fixed in Ref. \cite{Kekez:2003ri}.
These contributions are substantial, but if we take the chiral limit, 
all of them would tend to zero according to the GMOR relation, as 
required by the chiral symmetry of QCD. 
Thanks to this relation, even for the realistically large strange 
mass our approach has  
\begin{equation}
M_{s\bar s}^2 \approx 2 M_K^2 - M_\pi^2
\label{MssbarMKMpi}
\end{equation}
in a reasonably good approximation, at the 10\% level, whereby
\begin{equation}
M_{88}^2 \approx \frac{4}{3} M_K^2 - \frac{1}{3} M_\pi^2 \, .
\label{approxGMO}
\end{equation}
Equation (\ref{M88}) is thus revealed as a variant of the standard
$\eta_8$ Gell-Mann--Okubo relation (\ref{approxGMO}) featuring 
only pion and kaon masses, which are not affected by the anomaly 
{\footnote{The kaon is protected from mixing not only by 
isospin, but also by strangeness.}}.
This is not surprising, as the role of $M_{88}^2$ in 
Eq. (\ref{M2NA}) is the non-anomalous $\eta_8$ ``mass''.
Similarly, even $M_{00}$ (\ref{M00}), which by Eq. (\ref{MssbarMKMpi})
gives the ``$\eta_0$ Gell-Mann--Okubo relation"
\begin{equation}
M_{00}^2 \approx \frac{2}{3} M_K^2 + \frac{1}{3} M_\pi^2 \, ,
\label{approxGMOeta0}
\end{equation}
is just the
{\it non-anomalous part} of the $\eta_0$ ``mass" $M_{\eta_0}$. 
It however requires the anomalous, chiral-limit-nonvanishing
part to avoid the U$_{\rm A}$(1) problem.

In our DS approach, all the model masses $M_{q\bar q'}$ ($q, q' = u,d,s$) 
and corresponding $q\bar q'$ bound state amplitudes are
obtained in the ladder approximation. Thus, regardless of any 
concrete model, they are obtained with an interaction kernel
which cannot possibly capture the effects of gluon anomaly.
Fortunately, the large $N_c$ expansion indicates that the
leading approximation in that expansion describes the bulk
of main features of QCD.
The gluon anomaly is suppressed as $1/N_c$ and is viewed as 
a perturbation in the large $N_c$ expansion. It is thus a 
meaningful approximation \cite{Klabucar:1997zi} to consider 
the gluon anomaly effect only at the level of mass shifts and 
neglect its effects on the bound-state solutions. 

In the chiral limit and, as it will turn out, the SU(3) flavor limit, the 
gluon anomaly is coupled {\it only} to the singlet combination 
$\eta_0$ (\ref{eta0def}). Only the $\eta_0$ mass receives, from 
the gluon anomaly, a contribution which, unlike quasi-Goldstone 
masses $M_{q\bar q'}$'s comprising ${\hat M}^2_{NA}$, does {\it not} 
vanish in the chiral limit. 
As discussed in detail in 
Ref. \cite{Klabucar:1997zi},
in the present bound-state context it is most convenient to adopt 
the standard way (see, e.g., Refs. \cite{Miransky,Donoghue:dd})
to {\it parameterize} the anomaly effect. We thus break the 
U$_{\rm A}$(1) symmetry
by shifting the $\eta_0$
(squared) mass by an amount denoted by $3\beta \equiv \lambda_\eta$ 
(in the respective notations of Refs. \cite{Kekez:2000aw,Klabucar:2001gr}
and Ref. \cite{Klabucar:1997zi}). The complete mass matrix
is then ${\hat M}^2 = {\hat M}^2_{NA} +  {\hat M}^2_A$, where
\begin{equation}
{\hat M}^2_A = 
\left[ \begin{array}{ccl} 0 & \, 0 & \, 0 \\
                          0 & \, 0 & \, 0 \\
                          0 & \, 0 & 3\beta
      \end{array} \right]~.
\label{M2A}
\end{equation}
The value of the anomalous $\eta_0$ mass shift $3\beta$
is related to the topological susceptibility of the vacuum,
but in the present approach must be treated as a parameter to
be determined outside of our bound-state model, i.e., fixed by
phenomenology or taken from the lattice calculations such as Refs. 
\cite{Lucini:2004yh,DelDebbio:2004ns,Alles:2004vi,Alles:2005kx}.

We now want to incorporate the effects of the realistic breaking
of the SU(3) flavor symmetry into the description of the gluon 
anomaly. At this point it is customary to 
go straight to the nonstrange-strange ({\it NS--S}) basis
(\ref{etaNSdef})-(\ref{etaSdef}), but
before doing this, it is instructive to rewrite 
for a moment the matrix (\ref{M2A})  in the flavor, 
$|q\bar q\rangle$ basis, where {it has the pairing form,}
\begin{equation}
{\hat M}^2_A = \beta
\left[ \begin{array}{ccl} 1 & 1 & 1 \\
                          1 & 1 & 1 \\
                          1 & 1 & 1
        \end{array} \right]~,
\label{M2Aqq}
\end{equation}
since 
this may be the most transparent place to introduce the effect of flavor 
symmetry breaking into the anomalous mass shift. Namely, Eq. (\ref{M2Aqq}) 
tells us that due to the gluon
anomaly, there are transitions $|q\bar q\rangle \to |q' \bar q' \rangle $;
$q, q' = u, d, s$. However, the amplitudes for the transition from, and into,
light $u\bar u$ and $d\bar d$ pairs are expected to be different, namely 
larger, than those for the significantly more massive $s\bar s$.
To allow for the effects of the breaking of the SU(3) flavor symmetry,
we can write 
\begin{equation}
\langle q\bar q | {\hat M}^2_A |q' \bar q' \rangle =
b_q \, b_{q'} \, ,
\label{elementM2AqqX}
\end{equation}
where $b_q = \sqrt{\beta}$ for $q = u, d$ and $b_q = X \sqrt{\beta}$ for 
$q = s$. 
The anomalous mass matrix (\ref{M2Aqq}) is, in the flavor-broken case, 
thereby modified to
\begin{equation}
{\hat M}^2_A = \beta
\left[ \begin{array}{ccl} 1 & 1 & X \\
                          1 & 1 & X \\
                          X & X & X^2
        \end{array} \right]
 \to
\beta \left[ \begin{array}{ccl} 0 & 0 & \qquad \quad 0 \\
      0 & \frac{2}{3}(1-X)^2 & \frac{\sqrt{2}}{3}(2-X-X^2) \\
      0 & \frac{\sqrt{2}}{3}(2-X-X^2) & \quad \frac{1}{3}(2+X)^2
        \end{array} \right]
 \, ~,
\label{M2AqqX}
\end{equation}
where the arrow denotes rewriting ${\hat M}^2_A$ in the octet-singlet
basis (\ref{pi0def})--(\ref{eta0def}). Comparison with Eqs. (\ref{M2NA})
and (\ref{M2A}) shows that incorporating into the anomaly the flavor 
symmetry breaking, $X\neq 1$, leads to the following. First, the 
anomaly influences the $\eta_8 \leftrightarrow \eta_0$ transitions,
reducing the negative $M_{80}^2$ (\ref{M80}) by 
$\beta \, \frac{\sqrt{2}}{3}(2-X-X^2)$. More notably,
the $\eta_8$ and $\eta_0$ masses including {\it both} 
non-anomalous and anomalous contributions are given by 
Eqs. (\ref{M88}), (\ref{M00}), and (\ref{M2AqqX}) as
\begin{equation}
M_{\eta_8}^2 = M_{88}^2 + \frac{2}{3}(1-X)^2 \, \beta \, , 
\label{Meta8}
\end{equation}
\begin{equation}
M_{\eta_0}^2 = M_{00}^2 + \frac{1}{3}(2+X)^2 \, \beta \, .
\label{Meta0}
\end{equation}
Not only $M_{\eta_0}$ is modified, but the interplay of the 
gluon anomaly and flavor breaking modified the Gell-Mann--Okubo
relation (\ref{approxGMO}) as the anomaly becomes coupled also 
to $\eta_8$ and influences its mass $M_{\eta_8}$.

The Schwinger nonet formula, derived from the condition that the 
trace and determinant of 
${\hat M}^2 = {\hat M}^2_{NA} +  {\hat M}^2_A$
be equal to those of the same matrix in the basis of mass eigenstates, here
${\hat M}^2 = {\rm diag}(M_{\pi}^2, M_\eta^2, M_{\eta^\prime}^2)$,
now acquires the new term on the right-hand side:
\begin{eqnarray}
(4M_K^2-3M_\eta^2-M_{\pi}^2)
(3M_{\eta^\prime}^2+M_{\pi}^2-3M_K^2)
-
8(M_K^2-M_{\pi}^2)
\nonumber \\
= - 4 (M_K^2 - M_\pi^2) \, 3\beta \, (1-X^2) \,  ~,
\label{ourSchwingerForm}
\end{eqnarray}
were we also used Eq. (\ref{MssbarMKMpi}).
The usual Schwinger formula is known to be satisfied well 
for the vector and tensor nonets, but {\it not} for the 
pseudoscalar nonet \cite{Burakovsky:1998vc}.
Equation (\ref{ourSchwingerForm}) reduces to the usual Schwinger 
pseudoscalar-meson relation for the limit of no anomaly, 
$3\beta \to 0$, but also for just $X \to 1$, the
limit of no influence of the flavor symmetry breaking on the 
anomalous mass shifts. 
Thus, introducing only the anomalous shift of the $\eta_0$ mass 
still yields the usual Schwinger relation, as noted by Ref. 
\cite{Burakovsky:1998vc} in a different approach.

The pion remains decoupled from the etas as long as 
one stays in the isospin limit; i.e., after one adds 
the anomalous contribution ${\hat M}^2_A$ (\ref{M2AqqX}) 
to Eq. (\ref{M2NA}), one still can restrict 
oneself to $2\times 2$ submatrix in the subspace of etas. 
The calculationally convenient basis for that subspace
is the so-called {\it NS--S} basis:
        \begin{eqnarray}
        |\eta_\NSt\rangle
        &=&
        \frac{1}{\sqrt{2}} (|u\bar{u}\rangle + |d\bar{d}\rangle)
  = \frac{1}{\sqrt{3}} |\eta_8\rangle + \sqrt{\frac{2}{3}} |\eta_0\rangle~,
\label{etaNSdef}
        \\
        |\eta_\St\rangle
        &=&
            |s\bar{s}\rangle
  = - \sqrt{\frac{2}{3}} |\eta_8\rangle + \frac{1}{\sqrt{3}} |\eta_0\rangle~.
\label{etaSdef}
        \end{eqnarray}
In this basis, the $\eta$--$\eta^\prime$ mass matrix is
\begin{equation}
{\hat M}^2 =
             \pipsq{
                \begin{array}{ll}
         M_{\eta_{NS}}^2    &  M_{\eta_{S}\eta_{NS}}^2 \\
            M_{\eta_{NS}\eta_{S}}^2 &    M_{\eta_{S}}^2
                \end{array}
        }
     =
       \pipsq{
                \begin{array}{ll}
      M_{u\bar{u}}^2 + 2 \beta  & \quad \sqrt{2} \beta X \\
        \,  \sqrt{2} \beta X    & M_{s\bar{s}}^2 + \beta X^2
                \end{array}
        }
\begin{array}{c} \vspace{-2mm} \longrightarrow \\ \phi \end{array}
        \pipsq{
                \begin{array}{ll}
                        M_\eta^2        & \,\,  0 \\
                     \,   0               & M_{\eta'}^2
                \end{array}
        },
        \label{M2_NS-S}
\end{equation}
where the indicated diagonalization is given by
the {\it NS--S} mixing relations{\footnote{
The relation between the present approach and the two-mixing-angle 
scheme is clarified in the Appendix of Ref. \cite{Kekez:2000aw}.}}
\begin{equation}
|\eta\rangle = \cos\phi \, |\eta_\NSt\rangle
             - \sin\phi \, |\eta_\St\rangle~,
\,\,\,\,\,\,\,
|\eta^\prime\rangle = \sin\phi \, |\eta_\NSt\rangle
             + \cos\phi \, |\eta_\St\rangle~,
\label{eqno3}
\end{equation}
rotating $\eta_\NSt,\eta_\St$ to the mass eigenstates $\eta, \eta'$.
(In the last section we will use the effective-singlet-octet mixing 
angle $\theta$, defined by analogous mixing relations where 
$ \eta_\NSt \to \eta_8, \eta_\St \to \eta_0, \phi \to \theta$.
It is related to the completely equivalent {\it NS--S} mixing angle 
$\phi$ as $\theta = \phi - \arctan \sqrt{2} =  \phi - 54.74^\circ$.)

Now the {\it NS--S} mass matrix (\ref{M2_NS-S}) tells us that due to 
the gluon anomaly,
there are transitions $| \eta_\NSt \rangle \leftrightarrow | \eta_\St \rangle$.
As in the argument above Eq. (\ref{M2AqqX}), the amplitude for the transition 
from, and into, $\eta_\NSt$, need not be the same as those for the more massive $\eta_\St$.
The role of the flavor-symmetry-breaking factor $X$ is to allow for that
possibility. 
There are arguments \cite{Kekez:2000aw}, supported by
phenomenology, that the transition suppression is estimated well by
the nonstrange-to-strange ratio of respective quark constituent masses,
${\cal M}_u $ and ${\cal M}_s $.
Due to the Goldberger-Treiman relation, this ratio 
must be close \cite{Klabucar:1997zi,Kekez:2000aw} to the ratio 
of $\eta_\NSt$ and $\eta_\St$ pseudoscalar decay constants
$f_{\eta_\NSt}=f_\pi$ and $f_{\eta_\St}=f_{s\bar s}$.
In other words, we can estimate the flavor-symmetry-breaking suppression 
factor as $X\approx {\cal M}_u/{\cal M}_s$ or $X \approx f_\pi/f_{s\bar s}$.
(Yet another, but again closely related way of estimating $X$, 
is from the ratios of $\gamma\gamma$ amplitudes, as in 
Refs. \cite{Kekez:2000aw,Klabucar:2001gr}.)
In the present paper, we use $X\approx f_\pi/f_{s\bar s}$, because 
these decay constants are not only calculable in the DS approach, 
but also (in contrast to ``constituent quark masses") defined without 
any arbitrariness, and in the case of $f_\pi$ even experimentally
measurable. Our present model result $f_\pi/f_{s\bar s} = 0.6991$
(see Table \ref{piKssbarTable})
is reasonably close to $X_{\rm exp} \approx 0.78$ extracted
phenomenologically \cite{Kekez:2000aw} from the
{\it empirical} mass matrix ${\hat m}^2_{\rm exp}$ featuring 
experimental pion and kaon masses, or, after diagonalization, 
experimental $\eta$ and $\eta'$ masses -- see Eq. (\ref{m2exp}) below.

\subsection{ Mixing angle and other results
from \mbox{$\eta$--$\eta^\prime$} mass matrix  }
\label{mixAngleFromMassM}

In this subsection, let us first see what hints we get from phenomenology.
In our present notation, capital $M_a$'s denote the calculated, model
pseudoscalar masses, whereas lowercase $m_a$'s denote the corresponding
empirical masses. 

From our calculated, model mass matrix (\ref{M2_NS-S}) we form 
its empirical counterpart (\ref{m2exp})
by {\it i)} obvious
substitutions $M_{u\bar u} \equiv M_\pi \rightarrow m_\pi$ and
$M_{s\bar s} \rightarrow m_{s\bar s}$, and {\it ii)} by noting that
$m_{s\bar s}$, the ``empirical" mass of the unphysical $s\bar s$
pseudoscalar bound state, is given in terms of masses of physical
particles as $m_{s\bar s}^2 \approx 2 m_K^2 - m_\pi^2$ due to GMOR. 
Then,
\begin{equation}
{\hat m}^2_{\rm exp} =
        \pipsq{
                \begin{array}{ll}
                m_\pi^2 + 2 \beta       & \qquad \sqrt{2} \beta X \\
                        \sqrt{2}\beta X & 2 m_K^2 - m_\pi^2 + \beta X^2
                \end{array}
        }
\begin{array}{c} \vspace{-2mm} \longrightarrow \\ \phi_{\rm exp} \end{array}
        \pipsq{
                \begin{array}{ll}
                        m_\eta^2        & \, 0 \\
                        \, 0            & m_{\eta'}^2
                \end{array}
        } \, ,
        \label{m2exp}
\end{equation}
where the arrow indicates the diagonalization (\ref{eqno3})
for the angle value $\phi_{\rm exp}$.

Since $M_{u\bar u}$, obtained by solving the BS equation, is 
identical to our model pion mass $M_\pi$, it was fitted to the 
empirical pion mass $m_{\pi}$ in Ref. \cite{Kekez:2003ri}.
Similarly, $M_{u\bar s} \equiv M_K$ is fitted to the empirical 
kaon mass $m_K$, so that Eq. (\ref{MssbarMKMpi}) implies  
$M_{s\bar s}^2 \approx 2 m_K^2 - m_\pi^2$. 
We thus see that in our model mass matrix, the parts stemming
from its {\it non-anomalous} part ${\hat M}^2_{NA}$ (\ref{M2NA})
are already close to the corresponding parts in ${\hat m}^2_{\rm exp}$.
We can thus expect a good overall description of the masses in
$\eta$ and $\eta'$ complex. We now proceed to verify this expectation.

The {\it anomalous} entry $\beta$, along with $X$
(and then necessarily also $\phi$), is fixed phenomenologically 
if we require that they fit the masses $m_\eta$ 
and $m_{\eta'}$ in the empirical matrix (\ref{m2exp}) to their 
experimental values. This is achieved by requiring that trace and 
determinant of ${\hat m}^2_{\rm exp}$ have their experimental values, 
which leads (e.g., see \cite{Jones:1979ez,Scadron:1983jw}) to the relations 
\begin{equation}
\beta_{\rm exp} =
     \frac   { (m_{\eta'}^2 - m_\pi^2) (m_\eta^2 - m_\pi^2) }
           { 4 (m_K^2 - m_\pi^2) }  = 0.2785    \,\, {\rm GeV}^2 \, ,
\label{beta-izDetTr}
\end{equation}
\begin{equation}
X_{\rm exp} = \sqrt{2 \, 
\frac{(m_{\eta'}^2 - 2 m_K^2 + m_\pi^2)(2 m_K^2 - m_\pi^2 - m_\eta^2)}
     {(m_{\eta'}^2 - m_\pi^2)(m_\eta^2 - m_\pi^2)} } = 0.7791    \, ,
\label{X-izDetTr}
\end{equation}
\begin{equation}
\phi_{\rm exp} = \arctan
        \pib{
                \frac   {(m_{\eta'}^2 - 2m_K^2 + m_\pi^2) (m_\eta^2 -
                                 m_\pi^2)}
                                {(2m_K^2 - m_\pi^2 - m_\eta^2) (m_{\eta'}^2
                                 - m_\pi^2)}
        }^{1/2} = 41.88^\circ \, .
\label{phi-izDetTr}
\end{equation}

Now we want the analogous results from our theoretical mass matrix 
(\ref{M2_NS-S}), where only $\beta$ is not a calculated quantity. 
For example, in this subsection we use $X=f_\pi/f_{s\bar s} \approx 0.6991$
from Table \ref{piKssbarTable}.
We thus fix only $\beta$ by requiring that
the experimental value of the trace 
$m_\eta^2 + m_{\eta'}^2 \approx 1.22$ GeV$^2$ be fitted by 
the theoretical mass matrix (\ref{M2_NS-S}). This yields
\begin{equation}
\beta = \frac{1}{2+X^2} \, [ \, (m_\eta^2 + m_{\eta'}^2)_{\rm exp} -
                           (M_{u\bar u}^2 + M_{s\bar s}^2) \, ] \, ,
\label{fitTrace}
\end{equation}
whereby $X$, $M_{\pi^0}$, and $M_{s\bar s}$ from 
Table \ref{piKssbarTable} give us $\beta = 0.2723$ GeV$^2$,
in good agreement with $\beta_{\rm exp}$ (\ref{beta-izDetTr})
obtained from the empirical mass matrix (\ref{m2exp}).

Since $M_{u\bar u}^2 + M_{s\bar s}^2 = 2m_K^2$ holds to a very good
approximation due to GMOR, our approach satisfies well the first equality 
[from Eq. (\ref{fitTrace})] in
\begin{equation}
\beta \, (2 + X^2) = m_\eta^2 + m_{\eta'}^2 - 2 m_K^2 =
\frac{2 N_f}{f_\pi^2} \, \chi \, ,
\label{WittenVenez}
\end{equation}
where the second equality is the Witten-Veneziano (WV) formula \cite{WV}, 
with $\chi$ being the topological susceptibility of the pure Yang-Mills 
gauge theory. The WV formula with experimental masses and $f_\pi$ yields 
$\chi \approx (179 \,\, \rm MeV)^4$, which is in good agreement with our 
value $\chi = (177 \,\, \rm MeV)^4$, implied by Eq. (\ref{WittenVenez}) 
and our model values of $X=f_\pi/f_{s\bar s}$ and $\beta$ from 
Eq. (\ref{fitTrace}).
Our prediction for $\chi$ is also in reasonable agreement with the 
recent lattice results \cite{Lucini:2004yh,DelDebbio:2004ns,Alles:2004vi} 
considered in Ref. \cite{Alles:2005kx}. The central value of Lucini 
{\it et al.} $\chi_1 = (177 \pm 7 \,\, \rm MeV)^4$ \cite{Lucini:2004yh}
agrees precisely with our value, while their value obtained by a
different method, $\chi_2 = (184\pm 7 \,\, \rm MeV)^4$, is higher 
but still consistent with our $\chi$. However, 
$\chi_3 = (191  \pm 5  \, \rm MeV)^4$ 
of Ref. \cite{DelDebbio:2004ns} is too high for that.
On the other hand, 
our result is only marginally  too high to be consistent with 
the most precise lattice topological susceptibility so far,
$\chi_4 = (174.3 \pm 0.5 \pm 1.2^{+1.1}_{-0.2} \,\, \rm MeV)^4$ 
\cite{Alles:2004vi}.
In summary, our $\chi$ is consistent with 
\begin{equation}
\chi = (175.7 \pm 1.5 \, \rm MeV)^4 \,\, ,
\label{weightedAv}
\end{equation}
the weighted average of the recent lattice results
\cite{Lucini:2004yh,DelDebbio:2004ns,Alles:2004vi}.
Let us pause briefly to note that $\beta$ does not have to be 
treated as the parameter to be fixed by fitting the masses, since 
Eq. (\ref{WittenVenez}) enables one to determine $\beta$ from the 
lattice results on the topological susceptibility.  In fact, 
this is how we get the results in the column ${\cal C}$ of Table 
\ref{tab:eta-etap-mixing-OPE-new} in the last, concluding section. 

However, now we continue with $\beta$ from Eq. (\ref{fitTrace}) 
fitting $(m_\eta^2 + m_{\eta'}^2)_{\rm exp}$.  For this value, 
$\beta = 0.2723$ GeV$^2$, the values of
our {\it calculated} $\eta_{NS}$ and $\eta_S$ masses are  
\begin{equation}
M_{\eta_{NS}}^2 = M_{u\bar u}^2 + 2\beta = M_\pi^2 + 2\beta
= 0.5628 \,\, {\rm GeV}^2 = (750.2 \,\, {\rm MeV})^2
\label{MetaNS}
\end{equation}
and
\begin{equation}
M_{\eta_S}^2 = M_{s\bar s}^2 + \beta X^2 = 0.6545 \,\, {\rm GeV}^2
= (809.0 \,\, {\rm MeV})^2 \, .
\label{MetaS}
 \end{equation}
These are reasonable values, in good agreement with, {e.g.}, 
$\eta_{NS}$ and $\eta_S$ masses calculated in the dynamical SU(3) 
linear $\sigma$ model \cite{Klabucar:2001gr}.

The mixing angle is then determined to be $\phi = 40.17^\circ$ (or 
equivalently, $\theta = - 14.57^\circ$), for example through the 
convenient relation
\begin{equation}
\tan2 \phi = \frac{ 2 \sqrt{2} \beta X}{M_{\eta_S}^2-M_{\eta_{NS}}^2} \, .
\end{equation}
The diagonalization
of the $NS$-$S$ mass matrix gives us the $\eta$ and $\eta'$ masses:
\begin{eqnarray}
        M_{\eta}^2 &=& \cos^2 \phi ~M_{\eta_{NS}}^2    - 
\sqrt{2} \beta X  \sin 2\phi 
       +  \sin^2 \phi ~M_{\eta_{S}}^2 
        \label{eqno99a}
\\
        M_{\eta'}^2 &=& \sin^2 \phi ~M_{\eta_{NS}}^2   +
\sqrt{2} \beta X  \sin 2\phi             
       + \cos^2 \phi ~M_{\eta_{S}}^2~~.
        \label{eqno99b}
\end{eqnarray}
Plugging in the above predictions for $\beta, X, M_{\eta_{NS}}$, and 
$M_{\eta_S}$, our model $\eta$ and $\eta'$ masses then turn out to 
be $ M_{\eta} = 579$ MeV and $ M_{\eta'} = 939$ MeV, in good 
agreement with the respective empirical values of 547 MeV and 958 MeV.

The model values predicted in this and in the next subsection are
summarized in column ${\cal A}$ of Table \ref{tab:eta-etap-mixing-OPE-new}.
Columns ${\cal B}$ and ${\cal C}$ give analogously obtained results, but with 
differently chosen either flavor breaking parameter $X$ or the 
topologically susceptibility, i.e., $\beta$. The comparison of 
our predictions shows they are robust under these variations.

\subsection{Two-photon decays of $\eta$ and $\eta^\prime$}
\label{2photon}

Having obtained the predictions for the mixing in the $\eta$--$\eta'$
complex, we can get the predictions for the $\gamma\gamma$ decays of 
$\eta$ and $\eta'$ from the decay amplitudes 
$T_{\pi^0}^{\gamma\gamma}$ and $T_{s\bar s}^{\gamma\gamma}$
already given in Table \ref{piKssbarTable}. However, for the sake
of completeness, let us first briefly review how these 
amplitudes are obtained.

The transition between the neutral pseudoscalar meson $P$ and two photons 
$\gamma(k)$ and $\gamma(k')$ with momenta $k$ and $k'$ can be described by 
a scalar amplitude we denote $T_P(k^2,{k'}^2)$ \cite{Kekez:1998xr}. 
The special case of the decay $P\to\gamma\gamma$ into two real, 
on-shell photons corresponds to the $k^2={k'}^2=0$ amplitude
\begin{equation}
T_P^{\gamma\gamma} \equiv T_P(0,0) = {\rm const} \, , 
\label{P2gammaAmpl}
\end{equation}
so that by integrating over the phase space and summing
over the photon polarizations one gets the decay width
\begin{equation}
\Gamma(P\to\gamma\gamma) = \frac{\pi\alpha_{\rm em}^2 m_P^3}{4} \, 
|\, T_P^{\gamma\gamma} \,|^2 \, , \qquad (P = \pi^0, \eta, \eta') \, .
\label{P2gammaWidth}
\end{equation}

\begin{figure}
\includegraphics[height=57mm,angle=0]{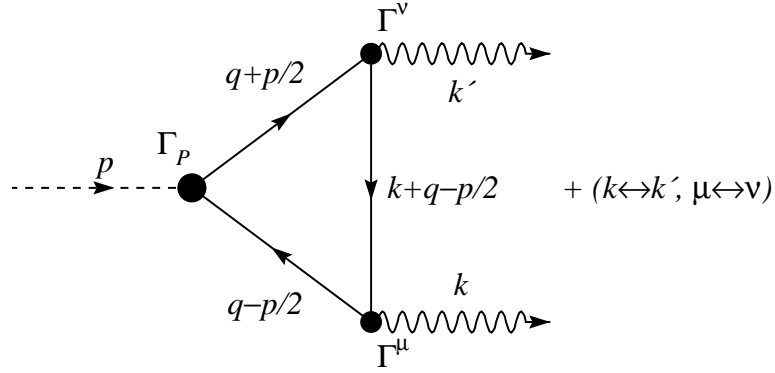}
\caption{The triangle graph (and its crossed graph) relevant for
the interaction of two photons of momenta $k$ and $k^\prime$ with
the neutral pseudoscalar meson $P$ of momentum $p$ represented by
the quark-antiquark bound-state vertex $\Gamma_P(q,p)$.
The quark-photon coupling is in general given by dressed vector
vertices $\Gamma_\mu(q_1,q_2)$, which in the free limit reduce to
$\hat{\cal Q} \gamma_\mu$, where
$\hat{\cal Q} = {\rm diag}({\cal Q}_u, {\cal Q}_d, {\cal Q}_s)
              = e  \, {\rm diag}(+2/3, -1/3, -1/3)$
is the flavor SU(3) quark charge matrix and $e$ is the
electromagnetic charge, $e=\sqrt{4\pi \alpha_{\rm em} } \,$.}
\label{fig:triangle}
\end{figure}

The calculation of the electromagnetic transition amplitudes
proceeds in the same way as in our earlier papers such as
Refs. \cite{Kekez:1996az,Klabucar:1997zi,Kekez:1998xr,Kekez:1998rw,Kekez:2001ph,Bistrovic:1999dy,Kekez:2000aw,Kekez:2003ri}, 
since the incorporation of the quark-photon
interactions is the same as adopted there through the scheme 
of a generalized impulse approximation, where all propagators, 
bound-state vertices, and quark-photon vertices are dressed.
(In the present application, this impulse approximation is 
illustrated by the $q\bar q$ pseudoscalar-$\gamma\gamma$ 
triangle graph in Fig. \ref{fig:triangle}.) They are all dressed 
mutually consistently, so that the pertinent Ward--Takahashi
identities are respected (e.g., see Refs. 
\cite{Roberts:1994hh,Bando:1993qy,Frank:1994gc}).
This is necessary for reproducing exactly and analytically 
anomalous $\gamma\gamma$ (on--shell) amplitudes{\footnote{And others, 
notably the ``box anomaly" process $\gamma \pi^+ \to \pi^+ \pi^0 $; see
Refs. \cite{Alkofer:1995jx,Bistrovic:1999dy,Bistrovic:1999yy,Cotanch:2003xv}.}
} 
in the chiral limit, and 
requires the usage of a dressed quark-photon vertex satisfying
the vector Ward--Takahashi identity. We employ the 
Ball--Chiu vertex~\cite{Ball:1980ay}, which is widely used (e.g., see Refs. 
\cite{Alkofer:2000wg,Roberts:2000aa,Maris:2003vk,Kekez:1996az,Klabucar:1997zi,Kekez:1998xr,Kekez:1998rw,Kekez:2001ph,Bistrovic:1999dy}
and references therein).

If one works in the {\it NS--S} basis 
(\ref{etaNSdef})-(\ref{etaSdef}), the diagonalization 
(\ref{eqno3}) and the amplitudes
\begin{eqnarray}
T_{\eta_{NS}}^{\gamma\gamma} &=& \frac{1}{\sqrt{2}} \,
 [\, T_{u\bar u}^{\gamma\gamma} + T_{d\bar d}^{\gamma\gamma}\, ]
=  \frac{5}{3} \, T_{\pi^0}^{\gamma\gamma} \, ,
\label{connectNSpi}
\\
T_{\eta_{S}}^{\gamma\gamma} &=& T_{s\bar s}^{\gamma\gamma} \, ,
\label{connectSssbar}
\end{eqnarray}
give the (mixing-dependent) amplitudes of the physical particles 
$\eta$ and $\eta'$:
        \begin{eqnarray}
        T_\eta^{\gamma\gamma}
        &=&
            \cos\phi\, \, \, T_{\eta_{NS}}^{\gamma\gamma} 
         - \sin\phi\, \, \, T_{\eta_S}^{\gamma\gamma}~,
\label{etaampl}
        \\
        T_{\eta^\prime}^{\gamma\gamma}
        &=&
            \sin\phi\, \, \, T_{\eta_{NS}}^{\gamma\gamma} 
         + \cos\phi\, \, \, T_{\eta_S}^{\gamma\gamma}~.    
\label{etaprimeampl}
\end{eqnarray}

In the preceding subsection, describing the breaking of 
the SU(3) flavor symmetry by $X=f_\pi/f_{s\bar s}$
led to $\phi = 40.17^\circ$. This mixing angle value
yields the physical amplitude values
\begin{equation}
  T_\eta^{\gamma\gamma} =   0.2706     \,\, {\rm GeV}^{-1} \, ,
 \qquad T_{\eta^\prime}^{\gamma\gamma} =  0.3410     \,\, {\rm GeV}^{-1} \, , 
\label{etasamplValues}
\end{equation}
which agree very well with the experimental amplitudes
\begin{equation}
(T_\eta^{\gamma\gamma})_{\rm exp} = 0.2724 \pm 0.0069 \,\, {\rm GeV}^{-1}\, ,   
\quad (T_{\eta^\prime}^{\gamma\gamma})_{\rm exp}  = 
  0.3417 \pm 0.0060  \,\, {\rm GeV}^{-1}\, .
\label{etasEXPamplValues}
\end{equation}
The corresponding calculated and experimental widths 
$\Gamma(P \to \gamma\gamma)$ ($ P = \eta,\eta'$) 
are respectively displayed in the columns ${\cal A}$ and ${\cal E}$ of 
Table \ref{tab:eta-etap-mixing-OPE-new}.
As already mentioned, other columns give our predictions 
based on somewhat different choices for either 
the flavor symmetry breaking parameter $X$ or the
topological susceptibility. Similarly to 
the results of the previous subsection, our results on
$\gamma\gamma$ decays turn out to be robust under these 
variations, since the resulting changes in the mixing angle 
are not excessive.

It is anyway very satisfying to note that the {\it mixing-independent} 
combination of our theoretical $\gamma\gamma$ decay amplitudes
\begin{equation}
| T_\eta^{\gamma\gamma} |^2 + | T_{\eta^\prime}^{\gamma\gamma} |^2 = 
| T_{\eta_{NS}}^{\gamma\gamma} |^2 + | T_{\eta_S}^{\gamma\gamma} |^2
  = 0.1895    \,\, {\rm GeV}^{-2}  \, 
\label{mixing-independentTheor}
\end{equation}
matches very well the corresponding experimental value
\begin{equation}
| (T_\eta^{\gamma\gamma})_{\rm exp} |^2 + 
| (T_{\eta^\prime}^{\gamma\gamma})_{\rm exp} |^2  = 
0.1909 \pm 0.0056 \,\, {\rm GeV}^{-2}\, ,
\label{mixing-independentEXP}
\end{equation}
because this match does not depend on the mixing angle at all.

\section{Discussion of results and conclusion}
\label{last}

\subsection{ On modification of analytic mass relations }
\label{AnoteOnGMO}

The interplay of the gluon anomaly and flavor symmetry breaking 
modifies some flavor SU(3) mass relations.  
The details of forming the mass matrix in subsection 
\ref{etaMasses} revealed the corrected $\eta_8$ Gell-Mann--Okubo 
relation (\ref{Meta8}), its $\eta_0$ analogue (\ref{Meta0}),
and Schwinger formula (\ref{ourSchwingerForm}). 
It is obvious that such results are independent of the DS approach
and even of any concrete dynamics, and depend only on the way 
the gluon anomaly and the SU(3) flavor symmetry breaking are 
implemented. It is thus not surprising that our Eqs. 
(\ref{Meta8}), (\ref{Meta0}), and (\ref{ourSchwingerForm}) are 
respectively equivalent to Eqs. (13), (15), and (38) of, e.g., 
Ref.~\cite{Li:2002hq} which does not address any dynamics.

Our primary aim concerning these mass relations
is to point out that all of them have been implicitly
contained in our DS approach to the $\eta$-$\eta'$ complex
and its results ever since Ref. \cite{Kekez:2000aw}  
incorporated into the DS approach
the effects of the flavor symmetry breaking on the anomalous mass shift.
However, it is instructive to consider these relations explicitly.
Already Eq. (\ref{ourSchwingerForm}) showed that the usual 
Schwinger formula, which is badly violated for pseudoscalars, 
acquires the correction proportional to $3\beta \, (1-X^2)$.
Note that the corrected Schwinger formula (\ref{ourSchwingerForm})
is identically satisfied if $\beta$ and $X$ are respectively 
given by Eqs. (\ref{beta-izDetTr}) and (\ref{X-izDetTr}), 
since these equations stem from the determinant and trace 
conditions, just like Eq. (\ref{ourSchwingerForm}) itself.
As for the $\eta_8$ Gell-Mann--Okubo formula, the original one 
(\ref{approxGMO}) was modified in Eq. (\ref{Meta8}) by the term 
\begin{equation}
\delta  \equiv \frac{2}{3} \, (1-X)^2 \, \beta  
        =  \chi \, \frac{4 N_f}{3 f_\pi^2} \, \frac{(1-X)^2}{2+X^2}~,
\label{corrGMO}
\end{equation}
where the last equality comes from expressing $\beta$ by the 
topological susceptibility $\chi$ through the WV relation 
(\ref{WittenVenez}).
This result is independent of
our concrete model, or even of the DS approach in general, as 
essentially the only strong simplifying assumption 
was the one of the (flavor-broken) nonet symmetry. 
If $X = f_\pi/f_{s\bar s}$ is chosen, the first-order 
flavor symmetry breaking estimate of $f_{s\bar s}$
(which is in our model satisfied better than 0.5\% 
-- see Table \ref{piKssbarTable}),
$f_{s\bar s} \approx 2 f_K - f_\pi$, 
permits expressing the correction (\ref{corrGMO})   
exclusively through experimental and lattice 
values. So, the experimental $f_\pi, f_K$ and 
the central value of the weighted average (\ref{weightedAv}) 
of the recent lattice results 
\cite{Lucini:2004yh,DelDebbio:2004ns,Alles:2004vi}
on $\chi$ gives $\delta \approx (127 \, {\rm MeV})^2$ independently 
of any model -- but in good agreement with our model result 
$\delta = (128 \, {\rm MeV})^2$ [from our model values of 
$X=f_\pi/f_{s\bar s}$ and $\beta$ from Eq. (\ref{fitTrace})].
Equation (\ref{Meta8}) then
yields $M_{\eta_8} = 608$ MeV, in excellent agreement 
with the chiral perturbation theory result 
of some 610 MeV \cite{Donoghue:1986wv}. 
It is satisfying that regardless of whether we take 
the lattice results for $\chi$ and empirical results for 
decay constants, or our model values for $\beta$ and $X$,
our correction (\ref{corrGMO}) is in reasonable agreement 
with Eq. (13.9) of Hagiwara et al. \cite{Hagiwara:2002fs}
(where our $\delta/M_{88}^2 = \Delta$ in their notation).
It gives $\theta = -14.1^\circ$,
in good agreement with our model results discussed in
the next, main subsection of the conclusion.

\subsection{ Model results }
\label{ModelResults}

The bulk of our results on the $\eta$--$\eta'$ complex are 
summarized in Table \ref{tab:eta-etap-mixing-OPE-new}, namely 
our model predictions for the masses, the flavor symmetry breaking 
parameter $X$, 
then $3\beta$, the anomalous squared mass of $\eta_0$ in 
the chiral or SU(3) flavor limit (where $\eta' = \eta_0$),
the mixing angle $\theta = \phi - \arctan \sqrt{2}$
and the $\eta, \eta' \to \gamma\gamma$ decays. 
The results in the column ${\cal A}$ were also given in the text
in the course of explaining our procedure. This column was 
obtained by assuming that the flavor symmetry breaking parameter 
$X$ is given by our model value of $f_\pi/f_{s\bar{s}}$.  The 
requirement (\ref{fitTrace}) that the inclusion of the anomalous 
mass contribution fit the sum of the (squared) experimental 
masses, $(m_\eta^2 + m_{\eta'}^2)_{\rm exp}$, then fixes the mass
matrix and the mixing angle which diagonalizes it. This
angle is then used to calculate the $\gamma\gamma$ decay
amplitudes and, equivalently, the corresponding widths.
The values in the column ${\cal B}$ were obtained in the same
way except that the flavor symmetry breaking parameter $X$
is estimated from the ratios of our calculated $\gamma\gamma$ 
amplitudes, $X = (T_{u\bar u}^{\gamma\gamma}/{\cal Q}_u^2)
               /(T_{s\bar s}^{\gamma\gamma}/{\cal Q}^2_s)$,
as explained in Ref. \cite{Kekez:2000aw}.
In the column ${\cal C}$ we again use $X=f_\pi/f_{s\bar{s}}$
as in the column ${\cal A}$, but instead of fixing $\beta$ by fitting
$(m_\eta^2 + m_{\eta'}^2)_{\rm exp}$,
we obtain $\beta$ through the WV relation (\ref{WittenVenez}) 
from the central value of the weighted average (\ref{weightedAv})
of the recent results on the topological susceptibility calculated 
on the lattice \cite{Lucini:2004yh,DelDebbio:2004ns,Alles:2004vi}.
Thus, the column ${\cal C}$ gives the masses, mixing and $\gamma\gamma$ 
decay widths without any parameter fitting whatsoever. In spite 
of this, the results in the column ${\cal C}$ are just as consistent 
with the experimental masses and decay widths as the results 
in the columns in which $\beta$ was used for fitting.

In the column ${\cal D}$ we give the best $\chi^2$-fit to the 
experiment which our theoretical $\gamma\gamma$ decay amplitudes 
$T_{\pi^0}^{\gamma\gamma}$ and $T_{s\bar s}^{\gamma\gamma}$ 
can give using the mixing angle as a free fitting parameter,
regardless of the results on the mixing angle from 
the $\eta$--$\eta'$ mass matrices. Nevertheless, 
the comparison with other columns in Table
\ref{tab:eta-etap-mixing-OPE-new} shows that what we find 
from $\eta, \eta^\prime \to \gamma\gamma$ processes is 
actually close to what we find from the mass matrix, which 
is of course very satisfying. Actually, the comparison of 
all results in Table \ref{tab:eta-etap-mixing-OPE-new} shows 
generally that our theoretical results in all columns are 
similar among themselves, exhibiting robustness under 
input variations,  and all are in reasonable agreement 
with the experimental results in column ${\cal E}$.

\begin{table}
\caption{Various theoretical results on $\eta$ and $\eta'$ and comparison 
with their experimental masses and $\gamma\gamma$ decay widths. All 
calculated quantities were obtained with
parameters which gave the very good description of pions and kaons in
Ref. \cite{Kekez:2003ri}, i.e., Eq. (\ref{StandardParameterSet-new}) and
${\widetilde m}_{u,d}=3.046$ MeV, ${\widetilde m}_s = 67.70 $ MeV.
The contents of the columns is this:
${\cal A}$) Summarized predictions of subsecs. \ref{etaMasses} and
\ref{mixAngleFromMassM}, i.e., with $X = f_\pi/f_{s\bar{s}}$.
${\cal B}$) $X$ estimated from $P\to\gamma\gamma$ {amplitudes} (see text).
In both ${\cal A}$ and ${\cal B}$, the parameter
$\beta$ is fixed by fitting the experimental value of
the mass matrix trace, Eq. (\ref{fitTrace}).
${\cal C}$) $X = f_\pi/f_{s\bar{s}}$ again, while $\beta$ is not a fitting 
parameter but obtained from $\chi = (175.7 \, {\rm MeV})^4$, the 
weighted average (\ref{weightedAv}) of the recent lattice 
topological susceptibilities
\cite{Lucini:2004yh,DelDebbio:2004ns,Alles:2004vi}.
The results of this column are thus obtained without any free parameters.
${\cal D}$) Starting from the calculated $\gamma\gamma$ amplitudes
$T_{\pi^0}^{\gamma\gamma}$ and $T_{s\bar s}^{\gamma\gamma}$
in Table \ref{piKssbarTable}, the mixing angle value is obtained as
the fitting parameter in the best $\chi^2$ fit to the experimental
amplitudes (\ref{etasEXPamplValues}) of $\eta$ and $\eta^\prime$.
${\cal E}$) experimental values.
The widths $\Gamma(\eta,\eta^\prime\to\gamma\gamma)$ are calculated using the
experimental masses. The $\gamma\gamma$ decay widths are in units of keV,
masses are in units of MeV, $3 \beta$ in units of MeV$^2$, while $X$ and
the mixing angles are dimensionless.
}
\begin{ruledtabular}
\begin{tabular}{cccccc}
                                     & ${\cal A}$            & ${\cal B}$           & ${\cal C}$          & ${\cal D}$                 &${\cal E}$                   \\
\hline
 $\theta$                            &{$-14.57^\circ$}       &{$-14.93^\circ$}      &{$-15.18^\circ$}	  & {$-14.27^\circ$}   &                    \\
 $M_\eta$                            &    {579.3  }          & {576.1  }            & 577.1               &                   & {$547.75\pm 0.12$} \\
 $M_{\eta^\prime}$                   &    {939.0  }          & {941.0  }            & 932.0               &                   & {$957.78\pm 0.14$} \\
 $X$                                 & 0.6991                & {0.7124  }           & 0.6991              &                   &                    \\
 $3\beta$                            & {816951}              & {810835}             & {798060}          &                   &                    \\
 $\Gamma(\eta\to\gamma\gamma)$       & {0.5034  }            & {0.5114  }           & {0.5170  }          & {0.4968  }        & {$0.510\pm 0.026$} \\
 $\Gamma(\eta^\prime\to\gamma\gamma)$& {4.272  }             & {4.229  }            & {4.199  }           & {4.308  }         & {$4.29\pm 0.15$}   \\
\end{tabular}
\end{ruledtabular}
\label{tab:eta-etap-mixing-OPE-new}
\end{table}

Let us now compare the present results with our earlier work 
\cite{Klabucar:1997zi,Kekez:2000aw} on the 
$\eta$--$\eta^\prime$ complex. It also employed the consistently 
coupled DS approach but using the Jain--Munczek effective 
interaction \cite{Jain:1993qh} instead of our Eq. (\ref{ourAlpha_eff}). 
The treatment of the $\eta$--$\eta^\prime$ complex used here 
was largely formulated already in Ref. \cite{Klabucar:1997zi},
except that it did not consider the interplay of the SU(3) flavor 
symmetry breaking and the gluon anomaly, which was taken into
account in Ref. \cite{Kekez:2000aw} and found important for the
successful meson phenomenology. The present paper, employing 
a different interaction, confirms its importance and further
clarifies it by displaying explicitly the interplay of $\beta$
and $X$ in the mass relations (\ref{Meta8})-(\ref{ourSchwingerForm}).  
While the reproduction of the $\eta$ and $\eta^\prime$ masses 
in Ref. \cite{Kekez:2000aw} was similarly successful, 
we find that our $\eta, \eta^\prime \to \gamma\gamma$ decay
widths agree with the experiment much better now, although 
the error bars shrunk as the precision of the experimental 
widths substantially increased in the meantime~\cite{Eidelman:2004wy}. 

Concerning the agreement with other approaches, we may 
point out that the $\eta$--$\eta'$ mass matrix obtained
on lattice by UKQCD collaboration \cite{McNeile:2000hf}
agrees reasonably well with our model mass matrix if we 
insert in it our values of $\beta$, $X$, $M_{u\bar u}$,
and $M_{s\bar s}$.
We should also recall that  already Ref. \cite{Kekez:2000aw} 
clearly showed that our DS approach and results are not in 
conflict, but in fact agree very well with results in the
two-mixing-angle scheme (reviewed and discussed in, {e.g}, 
Ref. \cite{Feldmann:1999uf}). Actually, Ref. \cite{Kekez:2000aw} 
showed that our results can also be given in the two-mixing-angle 
scheme, but it is defined with respect to the mixing of 
decay constants, and therefore in our case, as in the 
DS approach in general, the scheme with one angle defining 
the mixing of the states is more convenient.

Another important general feature of the consistent DS approach
which also holds in the present paper, is the correct chiral QCD 
behavior, so that all light pseudoscalar mesons constructed in our 
approach are both $q\bar q$ bound states and (almost-)Goldstone 
bosons of D$\chi$SB. In the chiral limit for all quark flavors,
the anomalous mass matrix ${\hat M}^2_{A}$ [with $X\to 1$, 
Eq. (\ref{M2A})] is
the only nonvanishing contribution as ${\hat M}^2_{NA} \to 0$. 
Thus, in this limit $\eta \to \eta_8$ with vanishing mass, and
$\eta' \to \eta_0$ with the mass $\sqrt{3\beta}$ (= 0.904 GeV in
column ${\cal A}$ and similarly in the other columns), which is the only 
non-vanishing pseudoscalar mass in that limit, being induced purely
by gluon anomaly. This chiral-limit mass $\sqrt{3\beta}$ is thus
only 6\% below the experimental $\eta'$ mass. Even with the 
realistic breaking of the chiral symmetry (as in Table 
\ref{tab:eta-etap-mixing-OPE-new}) which rises the $\eta$ by 
more than 0.5 GeV, our mixing angles show that $\eta$ and $\eta'$
are much closer to, respectively, the (almost-)Goldstone octet 
$\eta_8$ and non-Goldstone singlet $\eta_0$ than to $\eta_\NSt$ 
and $\eta_\St$ (which is why in this section we switched from 
$\phi$ to $\theta$).  
For example, column ${\cal A}$, with $\theta = -14.57^\circ$, implies
\begin{equation}
|\eta\rangle = 0.968 \, |\eta_8\rangle
             + 0.257 \, |\eta_0\rangle~,
\,\,\,\,\,\,\,
|\eta^\prime\rangle = - \, 0.257 \, |\eta_8\rangle
                      + 0.968 \, |\eta_0\rangle~, 
\end{equation}
where squaring of the coefficients $\cos\theta=0.968$ and 
$\sin\theta= - 0.257$ says that $\eta$ is 93.4\% $\eta_8$ 
and 6.6\% $\eta_0$, and reverse for $\eta^\prime$. 

In conclusion, the consistently coupled DS approach with
the effective interaction (\ref{ourAlpha_eff}) enhanced by 
the gluon condensates gives a very good description of the 
$\eta$--$\eta'$ complex. This was achieved along the lines 
formulated in Refs. \cite{Klabucar:1997zi,Kekez:2000aw},
where the consistently coupled DS approach was extended
by assuming the mass shift of the singlet $\eta_0$ due to 
the gluon anomaly. After Ref. \cite{Kekez:2003ri} found the 
model parameters for which the gluon-condensate-enhanced 
interaction (\ref{ourAlpha_eff}) leads to a sufficiently 
strong D$\chi$SB, pions and kaons as (quasi-)\-Gold\-sto\-ne 
bosons of QCD, and their successful DS phenomenology, 
this minimal extension was the only new element in the
otherwise fixed model. This was enough to successfully
model $\eta$ and $\eta'$ mesons without any parameter
re-fitting, which is not very surprising after the 
success with pions and kaons in Ref. \cite{Kekez:2003ri}. 
In Refs. \cite{Klabucar:1997zi} and especially \cite{Kekez:2000aw} 
we have already given good descriptions of $\eta$ and $\eta'$
employing the Jain--Munczek effective interaction \cite{Jain:1993qh}, 
which is however purely modeled at the low and intermediate energies.
In contrast, the important intermediate-momentum behavior of the 
presently used interaction (\ref{ourAlpha_eff}) may be actually 
understood in terms of gluon condensates, instead of just modeled. 
In addition, this interaction has presently given the description 
of $\eta$ and $\eta'$ which is on the whole somewhat better than 
in Refs. \cite{Klabucar:1997zi,Kekez:2000aw}, especially in view
of the increased precision of the $\eta,\eta' \to \gamma\gamma$
measurements \cite{Eidelman:2004wy}.

\vskip 3mm

\section*{Acknowledgment}
D. Klabu\v{c}ar acknowledges the partial support of Abdus Salam ICTP
at Trieste, where the largest part of this work was written.


\end{document}